# Negative-polarity nanosecond-pulsed cryogenic plasma in liquid nitrogen

Danil Dobrynin and Alexander Fridman

C&J Nyheim Plasma Institute, Drexel University, Philadelphia, PA, USA

**Abstract**

In this work we report the results of imaging and spectroscopic measurements of optical emission spectrum of negative nanosecond-pulsed cryogenic discharge in liquid nitrogen. With the application of lower electric fields, the discharge first ignites as a "faint glow" around the high voltage needle electrode, while when the applied electric field reaches transition value of around 5 MV/cm, the discharge mode switches to negative in-liquid streamer. Optical emission spectrum of the discharge is populated by the molecular nitrogen emission bands, and their analysis shows that the pressures and temperatures of the negative streamers in liquid nitrogen are at least of few tens of atmospheres and around 140-150 K. The results of the study demonstrate similarity of positive and neganive streamers in the cryogenic in-liquid plasma conditions.

1. Introduction

Low-energy nanosecond-pulsed discharges in liquids have been studied extensively in the last decade as one of the extreme cases of plasmas generated in high-density environment (see, for example, [1-21]. Typically, these discharges are ignited using positive few-ns long high voltage pulses applied to electrodes in pin-to-plane or pin-to-pin configuration placed in dielectric liquid, usually water or transformer oil. Due to short pulse duration, low energy input and the inertia of the liquid, formation of macroscopic voids is impossible, while high applied electric field in the vicinity of the high voltage needle provide conditions for formation of a luminous mm-scale streamers that are assumed to be initiated directly in liquid [1-21]. While the exact mechanisms of the initiation and propagation of such discharges are largely unknown, one of the proposed scenarios is related to the formation of positive streamers, or direct ionization of liquid [2]. In contrast to the positive discharges, the number of studies of the negative ns-pulsed plasma in liquid is limited, because generally these discharges are believed to be related to the field emission processes from the electrode and/or by the Townsend-like processes in the active negative corona regime [15, 20, 21]. Negative discharges in liquid water ignited using the high voltage pulses with the similar characteristics, are either small in size ("glow" region around the electrode with sizes on the order of few tens to a hundred of microns) [15, 21], or, alternatively, have similar size and emission pattern as positive counterparts [20].

While a ns-pulsed discharges in water have been extensively studied, cryogenic liquids present an opportunity to study these discharges in an arguably simpler environment – compared to water, liquid nitrogen, argon, and other cryogenic liquids have low dielectric permittivity constants (thus limiting effects of electrostriction and associated formation of nanovoids), while emission spectra contain molecular and atomic lines that are suitable for diagnostics (compared to plasmas in water that in the initial stages emits broadband continuum) [22, 23]. In our recent studies, we were able to register molecular nitrogen emission during the ignition and propagation of the positive discharge in liquid nitrogen, estimating neutrals temperatures and densities, as well as local electric field values that indicate that the discharge develops via streamer ("electronic") mechanism [22, 24]. In this manuscript, we report experimental results of imaging and spectroscopic measurement of the emission spectrum of the discharge in liquid nitrogen ignited at a tip of a negative needle electrode. We compare these observations with those obtained for the positive-polarity nanosecond-pulsed cryogenic in-liquid nitrogen plasma, and claim similarity of positive and neganive streamers in the cryogenic in-liquid plasma conditions.

2. Materials and methods

To study negative nanosecond-pulsed discharge in liquid nitrogen (LN, Airgas, impurities ≤5 ppm of $O_2$ and ≤10 ppm of $CO_2$) we have used grounded copper needle electrode with the radius of curvature of 100 µm. Opposite disc electrode with the diameter of 2 cm, also made of copper, was placed at a distance of 1.5-4 mm from the needle and powered by FPG 120-01NM10 high voltage power supply (FID technology). The power supply generates 8 ns-long (at 90% of amplitude) pulses with typical rise time of about 1 ns (10% to 90% amplitude) and amplitude between 40 and 120 kV. The electrodes were placed inside of a double-walled borosilicate glass with in-house characterized light transmission in 300 – 800 nm range. For comparison of the channel size, positive discharge was ignited in the same setup, however in this case the needle and planar electrodes were swapped.

Using 4Picos intensified charge-coupled device (ICCD) camera from Stanford Computer Optics, USA, we have performed emitted light and shadow imaging of the discharge, as well as recorded its emission spectra. For that, the camera was equipped with VZM™ 450 zoom imaging lens that provides magnification of 0.75 – 4.5X. We have used AFG-3252 function generator (Tektronix, USA) to synchronize the discharge imaging camera with the power supply. For shadow imaging, a 75 W Xe arc lamp (6251NS, Newport, USA) was used as a source of back light. To record the spectra of the discharge emission in the visible range, a Princeton Instruments-Acton Research, TriVista TR555 spectrometer system operating at a single stage with 750 l/mm grating via 1 m single-leg fiber optic bundle with nineteen 200 µm fibers (190 – 1100 nm, Princeton Instruments, USA) was used. The system light transmission function in the range of interest was measured using quartz tungsten halogen calibrated lamp (63350, Newport, USA) and applied to correct the recorded discharge spectra. All measurements we done in a single-shot single accumulation regime, with the exception of the spectra recording which required 100 accumulations of the discharge operating at 5 Hz repetition frequency.

3. Results and discussion

Previously, negative nanosecond-pulsed discharges were studied for water plasmas – see, for example, [15] and [21]. Direct comparison of the discharge ignited by either positive or negative polarity high voltage pulses was done by Grosse et al. [15], who used a pin-to-pin electrode configuration (although sharpness of the 50 µm thick tungsten wire electrodes was not indicated, from the Figure 5 we estimate it to be of about 5 µm). They show that with the pulses of 20 kV (pulse duration of 10 ns and rise time of 2-3 ns) applied to the electrodes separated by 10 mm gap, the discharge appears to be of very similar size (~0.25 mm) and shape, regardless of the applied voltage polarity. In contrast, Seepersad et al. [21], for similar pulses (20 kV, 10 ns duration and 3 ns rise time) applied to pin electrode of 25 µm tip curvature with planar electrode placed at 1.5 mm distance recorded strikingly different discharge images: while in the positive mode had maximum size of ~0.15 mm, in the negative mode plasma appeared as a faint glow around the electrode. While these results seem contradictory, it is important to note that in the case of Grosse et al. [15] the applied maximum electric field (which we estimate to be ~25 MV/cm) was about 5 times higher than in the experiments done by Seepersad et al. [21] We attribute this to the streamer formation process that is governed by the Meek's criterion: if the applied electric field is below breakdown, discharges would mostly not start because avalanches do not generate high enough local electric field to transition to streamer [25]. In our experiments with generation of the negative discharge in liquid nitrogen, we have observed this transition: as shown in Figure 1, there is clear change of the discharge appearance with the applied voltage. Relatively low amplitude applied voltage pulses of 50 kV and less result in the discharge that appears as a faint "glow" around the needle electrode, while higher voltages produce streamer-like filaments with typical sizes of about 0.5 – 1 mm. Keeping in mind that typical size of this "glow" is close to the electrode tip curvature, we can conclude that relevant plasma processes are not determined by the non-local very energetic field-emission electrons, but rather by the Townsend-like processes in the active negative corona.

The observed negative streamers propagate towards the anode due to the space charge created by the electrons that move away from the cathode in more or less uniform manner, resulting in decrease of the electric field at the front [25]. Compared to the positive streamers that follow "focused" growth pattern due to mostly stationary positive charges, negative streamers, therefore, tend to start at higher voltages, and propagate to shorter distances [25], as seen also in Figure 2. For the same reasons, we observe slight increase of the breakdown electric field for the negative discharge for larger gaps. Generally, the observed negative in-liquid cryogenic streamers have similar "focused" growth pattern (lightning-like shape) as those observed for the positive in-liquid cryogenic streamers. It can be interpreted considering that according to the Meek's criterion the streamer length (about $20/\alpha$ for both positive and negative streamers, where $\alpha$ is the first Townsend coefficient) exceeds its radius (usually $1/\alpha$) about 20 times [26].

Two ns exposure time-resolved images of the single-shot discharge ignited in liquid nitrogen in streamer mode are shown in Figure 3. In line with other studies, the discharge is first detected at the front edge of the applied voltage pulse, followed by the "dark phase" and the growth of the emitting region on the falling edge of the pulse. Compared to positive discharge, propagation velocity of the negative mode is slightly lower: ~0.37±0.1 mm/ns (370±10 km/s) for negative discharge versus about 0.5 mm/ns for positive one [22], as typical for gas-phase streamers [25]. Negative discharge also lasts about 10 ns longer compared to the positive.

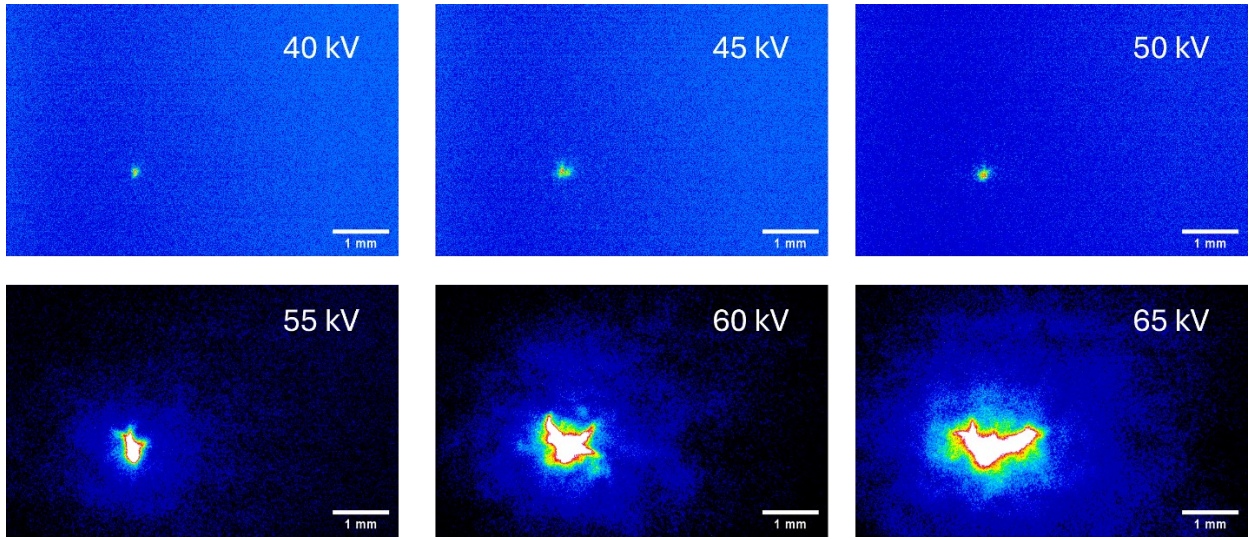

Figure 1. Integral images of the negative discharge in liquid nitrogen ignited with pulses of various amplitude. Exposure time 50 ns, discharge gap 2.25 mm.

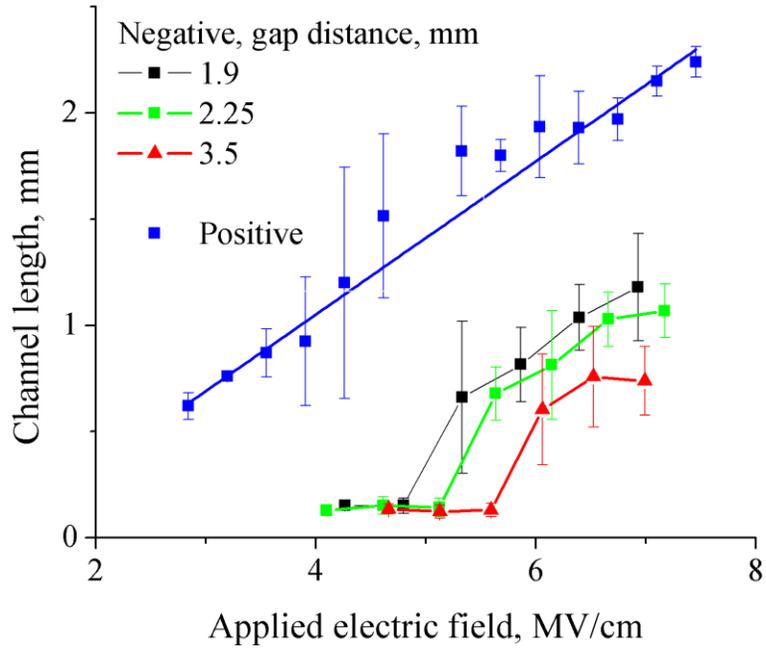

Figure 2. Comparison of the maximum channel length of the negative discharge in liquid nitrogen as a function of the applied electric field to the positive discharge.

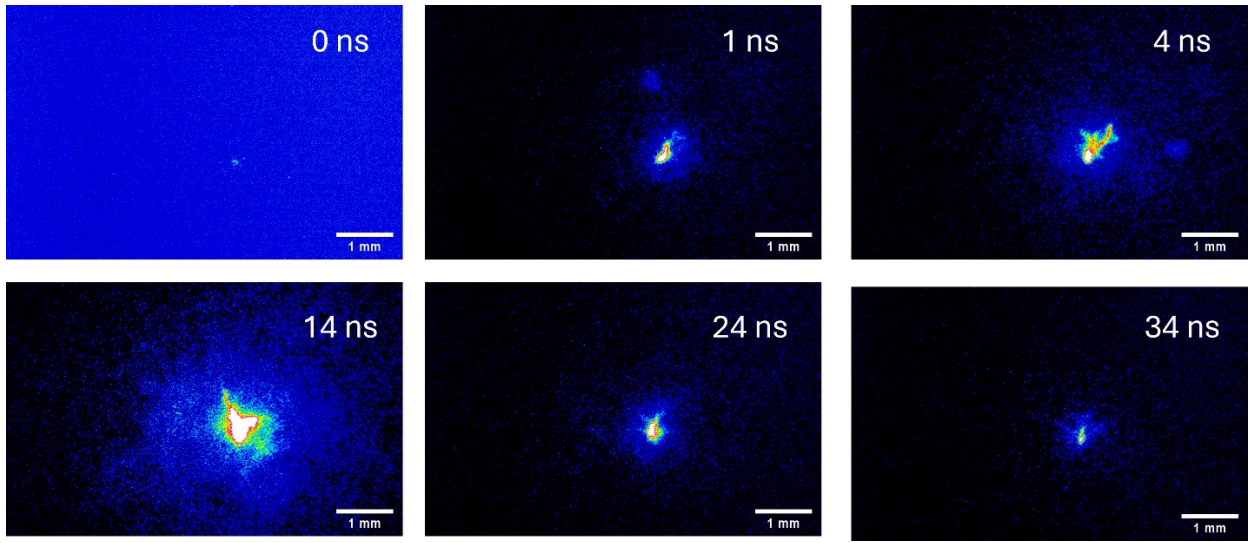

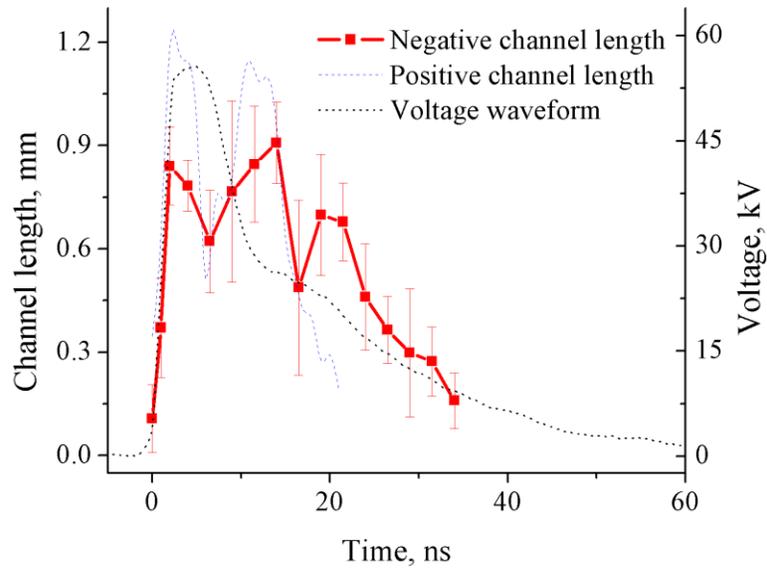

Figure 3. Time-resolved images of the negative discharge in streamer mode (2 ns exposure time, single accumulation, discharge gap 2.25 mm, 60 kV), and evolution of the negative discharge size in time compared to the positive one (60 kV, data reproduced from [22]).

Shadow imaging results of the discharge are shown in Figure 4. The images were taken using single accumulation with 50 ns exposure time. After the luminous phase of the discharge, gaseous areas roughly corresponding to the discharge size become apparent. In the later time points, these gaseous structures grow and eventually, around 50-100 µs, become spherical bubbles. For the "glow" mode, gaseous voids and bubbles appear smaller, likely due to lower energy deposition compared to stronger streamer discharge at higher voltages.

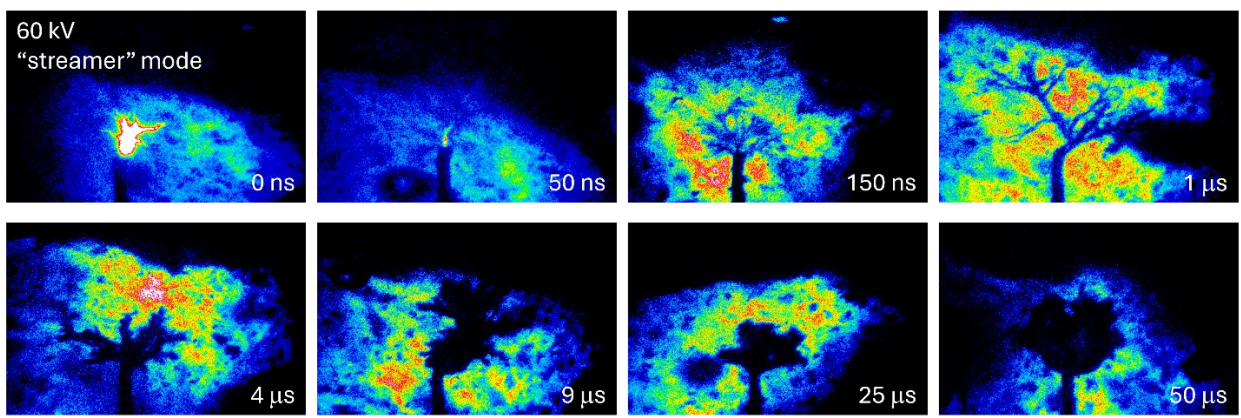

Figure 4. Shadow imaging of the negative discharge in liquid nitrogen – exposure time 50 ns, single accumulation, discharge gap 2.25 mm.

Discharge emission spectrum in the streamer mode was recorded using 50 ns exposure time and 100 accumulations of the discharge operating at 5 Hz. In the Figure 5, the recorded spectrum (corrected for the camera detector spectral sensitivity, transmission of optical components, and absorption by the liquid nitrogen) is shown in comparison with the positive discharge reproduced from [22]. Here, as in the case of the positive discharge in liquid nitrogen, emission spectrum contains molecular bands from the first and second positive systems of nitrogen, which increase in the intensity towards the longer wavelengths. Continuum emission in the 600 – 800 nm range that is present in the positive discharge spectrum, was not detected in the negative streamer mode, which could indicate the importance of ion–atom recombination and radiative charge exchange in the positive streamers [27]. Absence of strong continuum emission for the negative discharge was also observed in the water experiments reported in [15] and attributed to the role of field emission from the electrode that dominates over field ionization compared to the positive polarity. Specair simulation tool includes relevant broadening mechanisms, including Van der Waals broadening [28 – 31], which can be used for evaluation of the pressure in the discharge zone. Using values of pressure similar to the ones estimated in [24] for the positive streamer discharge in liquid nitrogen – about 50 atm, – we have fitted the measured significantly broadened molecular nitrogen bands to gauge the discharge temperature. The results obtained from the integral emission spectra (Figure 5) indicate temperatures in the 140-150 K range and are in a good agreement with the values obtained for the positive streamer plasma in liquid nitrogen [22-24].

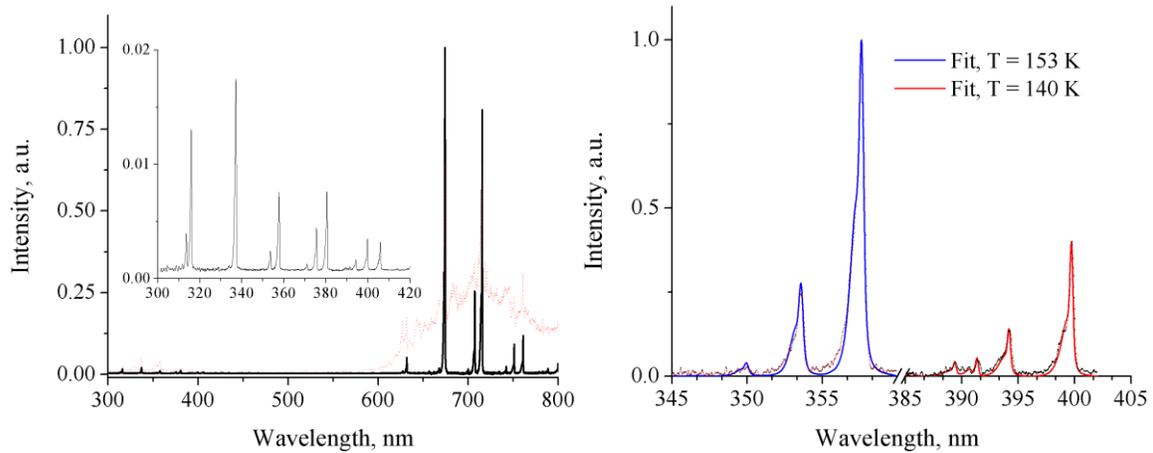

Figure 5. Left: emission spectrum of the negative discharge in liquid nitrogen (solid line) compared to the positive discharge (dotted line, data reproduced from [22]). Right: Specair-fitted (solid line) of the measured (dotted line)

4. **Conclusions**

In this manuscript we report on observation of negative nanosecond-pulsed cryogenic discharge in liquid nitrogen. We show that the discharge first ignites at lower applied electric fields as a "faint glow" around the high voltage needle electrode with typical sizes on the order of a hundred microns. This "glow" is probably sustained not by the non-local very energetic field-emission electrons, but rather by the Townsend-like processes in the active negative corona zone. When the applied electric field reaches transition value of around 5 MV/cm, the discharge mode switches to negative in-liquid streamer with the appearance similar to that of a positive in-liquid discharge. The negative cryogenic in-liquid streamers have similar "focused" growth pattern as those of positive cryogenic in-liquid streamers with the length-to-radius ratio about 20, in accordance with the Meek's criterion.

While shadow imaging demonstrates generation of gaseous voids following the discharge disappearance, initial high propagation velocity of the negative streamer suggest that the discharge is initiated directly in the liquid phase. Optical emission spectrum of the negative streamer discharge, in general, appears to be similar to the one of the positive streamer mode, and is populated by the molecular nitrogen emission bands. Using Specair simulation tool, we estimate that the pressures and temperatures of the negative streamers in liquid nitrogen are on the same order of magnitude as for the positive ones: few tens of atmospheres and around 140-150 K.

**Acknowledgement**
This work was supported by the US National Science Foundation award # 2108117, PI: Dobrynin.